\documentclass[referee]{aa} 
\usepackage{graphicx}
\newcommand{\1}{\,{\sc i}}
\newcommand{\2}{\,{\sc ii}}
\newcommand{\3}{\,{\sc iii}}

\usepackage{epsf}
\usepackage{amssymb}

\begin{document}
   \title{Phosphorus in the Diffuse Interstellar Medium}
   \author{V. Lebouteiller\inst{1}, Kuassivi\inst{2}, \&\ R. Ferlet\inst{1}}

   \offprints{V. Lebouteiller}

   \institute{1: Institut d'Astrophysique de Paris,
UMR7095 CNRS, Universit{\' e} Pierre \&\ Marie Curie, 98 bis boulevard Arago, 75014 Paris\\
              2: AZimov association, 14 rue Roger Moutte, 83270 St-Cyr sur Mer, France
 \\
   \email{leboutei@iap.fr}}

\date{Received; accepted 07/18/05}

   \abstract{
   We present \textit{FUSE} and HST/\textit{STIS} measurements of the P\2\
   column density toward Galactic stars. We analyzed P\2\ through the profile fitting of
   the unsaturated $\lambda$1125 and $\lambda$1533 lines and derived column densities integrated along the
   sightlines as well as in individual resolved components. We find that phosphorus is
   not depleted along those sightlines sampling the diffuse neutral
   gas. We also investigate the correlation existing between P\2\ and O\1\ column densities
   and find that there is no differential depletion between these two specie.
   Furthermore, the ratio $N$(P\2)/$N$(O\1) is consistent with the solar P/O value, implying that P\2\
   and O\1\ coexist in the same gaseous phase and are likely to evolve in parallel
   since the time they are produced in stars. We argue that phosphorus, as traced by
   P\2, is an excellent neutral oxygen tracer in various physical environments, except when ionization
    corrections are a significant issue. Hence, P\2\
   lines (observable with \textit{FUSE}, HST/\textit{STIS}, or with VLT/\textit{UVES} for the
   QSO sightlines) reveal particularly useful as a proxy for O\1\ lines when these are saturated or blended.

   \keywords{ISM: abundances, atoms, clouds, Galaxy: abundances, Ultraviolet: ISM}
   }

   \titlerunning{Phosphorus in the diffuse ISM}
   \authorrunning{Lebouteiller, Kuassivi \& Ferlet}

   \maketitle
%

\section{Introduction}

Phosphorus ($^{31}_{31}$P) is an odd-Z element which is thought to be mainly produced in
the same massive stars that form $\alpha$-elements (O, Ne, Mg, Si, S, Ar, ...). Although
the nucleosynthesis of odd-Z elements is still not well understood, phosphorus seems to
be produced during the carbon and neon burning in a hydrostatic shell (Arnett 1996).
Woosley \& Weaver (1995) found that no significant amount is expected to be synthesized
during the explosion phases and that the P yields should be metallicity dependent because
of the odd--even effect.

In the Galactic and extra--galactic diffuse interstellar medium (ISM), the phosphorus
gaseous phase abundance is largely unknown. Apart from the early investigations with the
\textit{Copernicus} satellite within the solar neighborhood (Jenkins et al. 1986; Dufton
et al. 1986), quite a few measurements are available in the distant ISM and almost none
in the extragalactic ISM. Nevertheless, the ISM phosphorus abundance provides an
important constraint for the dust/gas chemistry.

Dust grains are subject to dramatic chemical, morphological, and structural changes
during their life time, from the condensation of dusty cores within the outflows of
evolved stars to the final destruction in shocks or the formation of new generations of
stars and planetary systems. They undergo several growth (mantle accretion and molecules
adsorption) and erosion (photoprocessing, sputtering) periods as they move within dense
or diffuse media (Snow \&\ Meyers 1979; Seab 1987; Turner et al. 1991; O'Donnell \&\
Mathis 1997). In that respect, refractory elements are expected to be the most depleted
in the gas phase. On the contrary, C, N, and O are less depleted onto dust grains because
of their lower condensation temperatures (Field 1974; Lodders 2003). Hence, one third of
the oxygen is bound to rocky elements at most (Cardelli et al. 1997). With respect to its
relatively high condensation temperature, phosphorus can be expected \textit{a priori} to
be more heavily depleted.

Jenkins et al. (1986), in their ISM survey conducted with \textit{Copernicus} toward
about 80 stars, found through the analysis of the far-UV absorption lines of the dominant
ion stage P\2\ that phosphorus is not depleted along sightlines containing predominantly
warm low density neutral gas ($\lesssim$0.1~dex after updating the solar abundances from
Asplund et al. 2004) and is depleted by $\lesssim$0.5~dex in cooler and denser clouds.
Using the same dataset but with a different oscillator strength for the P\2\
$\lambda$1302 transition, Dufton et al. (1986) derived phosphorus abundances
systematically larger by $\approx0.2$~dex so that the previous findings still hold.
Finally, it must be added that in the cold diffuse interstellar cloud toward $\zeta$~Oph,
the depletion of phosphorus is $0.5\pm0.2$~dex as compared with $0.4\pm0.1$~dex for
oxygen, while in the warm diffuse cloud along the same sightline, P is depleted by
$0.2\pm0.1$~dex as compared with $0.0\pm0.3$~dex for O (Savage \&\ Sembach 1996).

The first detection of the PN molecule by Turner \&\ Bally (1987) through the 140~GHz
($J=3-2$) and 234~GHz ($J=5-4$) emission lines paved the way to extensive studies of the
abundance of phosphorus in molecular clouds. It was soon recognized that the transfer of
phosphorus from the gas to the solid and back to the gas phase was largely involving
carbon atoms via the HCP linear molecule (Turner et al. 1990). As a consequence,
phosphorus is believed to mainly reside in adsorbed HCP molecules which are then released
in the gas by photodesorption in warm media and readily photodissociated. This scenario
would account for the lack of depletion in the warm phase.

Since the advent of the Far Ultraviolet Spectroscopic Explorer (\textit{FUSE}) satellite
(Moos et al. 2000) and the \textit{STIS} instrument onboard HST, it is now possible to
probe denser clouds and investigate longer sightlines. We thus revisit and extend
previous works on interstellar phosphorus abundance by presenting new P\2\ measurements
toward Galactic sightlines obtained from \textit{FUSE} and HST/\textit{STIS} data.
Furthermore, in order to point out the relative behavior of phosphorus as compared with
$\alpha$-elements and possibly reveal a global trend toward the differential depletion
under many different physical conditions, we compare the phosphorus gaseous phase
abundance with that of oxygen. Oxygen abundance in the diffuse ISM has been extensively
studied over the last years (see e.g., Jensen et al. 2005, Cartledge et al. 2001,
Cartledge et al. 2004, and Andr\'e et al. 2003 $-$ hereafter A2003) and is used here as a
reference element. The observations and data analysis are described in Sections~2 and~3.
We present the results in Sect.~4. Final conclusions are given in Sect.~5.

\section{Observations}\label{sec:obs}

We have selected 10 sightlines toward Galactic stars with distances up to $\approx 5$~kpc
in order to scan the distant ISM. Properties of the targets and the sightlines are
summarized in Table~\ref{tab:prop}. Tables \ref{tab:fuseobs} and \ref{tab:stisobs}
provide the log of the \textit{FUSE} and HST/\textit{STIS} observations, respectively.
Given the Galactic latitudes and distances of the stars, all the sightlines intersect
clouds in the Galactic disk, except the sightline toward HD121968 which possibly can
intersect clouds in the halo.

All the \textit{FUSE} spectra were obtained through the large
30\arcsec$\times$30\arcsec~(LWRS) aperture which results in a resolving power $R\equiv
\lambda/\Delta \lambda \approx 20,000$~(or $\Delta v \approx$ 15 km s$^{-1}$, FWHM). This
spectral resolution depends on the co--addition procedure used to reconstruct the total
exposure and varies with the wavelength and the detector. Hence, we did not attempt to
co--add different detectors, in order to minimize both the distortion of the resulting
Point Spread Function (PSF) and the propagation of the Fixed Pattern Noise (FPN) proper
to each detector. The detailed reduction, calibration and co--addition procedures can be
found in A2003 who reduced most of the present data in order to study the neutral oxygen
and hydrogen content along the sightlines.

The \textit{STIS} observations were taken with the far-ultraviolet MAMA detector equipped
with the E140H grating. However, three different apertures were used. The
$0\farcs1\times0\farcs03$~aperture provides a resolving power $R\approx 200,000$~(or
$\Delta v \approx 1.5$~km~s$^{-1}$, FWHM). The two others, $0\farcs2\times0\farcs09$ and
$0\farcs2\times0\farcs2$, provide a spectral resolution of $R\approx 110,000$ (velocity
resolution of $\Delta v \approx 2.7$~km~s$^{-1}$, FWHM). Again, details about the data
reduction can be found in A2003.

Among the 10 targets we present, 8 were analyzed by A2003 with a particular concern on
the O\1\ column density. The other targets of their sample were not observed with
\textit{STIS} at the wavelength of the P\2\ $\lambda$1533 line (see next section).
Furthermore, we analyzed \textit{STIS} spectra of two additional targets, HD24534 and
HD121868.

Finally, we also compiled published P\2\ and O\1\ measurements in the Milky Way and along
a few extragalactic sightlines. These are listed in Table~\ref{tab:lit} and discussed in
Sect.~\ref{sec:others}.

\section{Data analysis}\label{sec:data}

Because of the ionization potential of O\1\ (13.62~eV as compared with 13.60~eV for H\1)
and the efficient charge exchange between O\2\ and H\1, O\1\ is expected to be the
dominant ionization state of oxygen in the diffuse neutral gas, and thus a good tracer of
the neutral gas. Observations of A2003 confirm this finding. On the other hand, the
ionization potentials of P\1\ and P\2\ (resp. 10.49~eV and 19.72~eV) suggest \textit{a
priori} that P\2\ should be the dominant state of phosphorus in this gaseous phase.
However a fraction of P\2\ atoms could actually reside in a potential ionized gaseous
phase where oxygen is into O\2\ and hydrogen into H\2. The fraction is unknown and
depends on the ionizing radiations illuminating the diffuse clouds. The present study
will help to identify this possible correction.

Absorption lines were analyzed assuming Voigt profiles, by using the profile fitting
program \texttt{Owens}. This fortran code, developed by M.~Lemoine and the \textit{FUSE}
French team, is particularly suitable for simultaneous fits of far-UV spectra (Lemoine et
al. 2002). A great advantage of this routine is the ability to fit different spectral
domains and various species in a single run. It was then possible to analyze
simultaneously the P\2\ and O\1\ absorption lines, together with Cl\1, C\1, and S\1\
lines. The use of these species in a simultaneous fit allowed us to check and constrain
the radial velocity structure of the sightlines when uncertain. The P\2\ and O\1\ lines
we analyzed being unsaturated, the column density determination does not depend on the
$b$-parameter. The errors on the column densities are calculated using the $\Delta
\chi^2$ method described in H{\'e}brard et al. (2002) and include the uncertainties on
all the free parameters such as the continuum shape and position. All the errors we
report are within 1~$\sigma$.

Numerous O\1\ lines are observed in the HST/\textit{STIS} $+$ \textit{FUSE} spectral
ranges, with oscillator strengths ($f$) spanning several orders of magnitude. However,
the main constraint on the O\1\ column density consists in using the weak 1355.5977 \AA\
intersystem transition ($f = 0.116\times10^{-5}$). The total O\1\ column densities toward
most of the present sightlines were derived by A2003 using this line. However, we have
updated these values by deriving column densities of each cloud along the sightlines and
by performing a simultaneous analysis of O\1\ and P\2\ lines.

A total of seven P\2\ lines are available in the far-UV spectral domain. Thanks to the
combination of datasets, the phosphorus atoms content is readily explored through
transitions spanning more than 2 orders of magnitude in oscillator strengths, always
allowing the choice of \textit{adequat} lines for a particular study. Wavelengths and
$f$-values are from the revised compilation of atomic data by Morton (1991; 2003). One
must be aware that oscillator strengths of P\2\ lines could be relatively uncertain.
Indeed, phosphorus atomic data have been poorly investigated and no laboratory
experiments exist.

In the present study, the three P\2\ lines at 961.0412~\AA~($f = 0.349\times10^{0}$),
963.8005~\AA~($f = 0.146\times10^{1}$) and 972.7791~\AA~($f = 0.210\times10^{-1}$),
observable with \textit{FUSE}, are located in a region overcrowded with strong absorption
lines $-$ mainly H\1\ lines from the Lyman serie and H$_2$~lines. We thus rejected these
transitions because blended. In addition, the two first lines are heavily saturated, thus
preventing reliable P\2\ column determinations.

Similarly, we avoided the P\2\ lines at 1152.8180~\AA\ ($f = 0.245\times10^{0}$,
observable with \textit{FUSE}), and at 1301.8743~\AA\ ($f = 0.127\times10^{-1}$,
observable with \textit{STIS}) since they both present most of the time saturation
effects. It must be added that the later is most often blended with the extremely strong
O\1\ line at 1302.1685~\AA, even at the highest \textit{STIS}~resolution.

The P\2\ line at 1124.9452~\AA\ ($f = 0.248\times10^{-2}$, observable with \textit{FUSE})
is the weakest transition available and is always found to lie on the linear part of the
curve of growth. When detected, it can provide an accurate measurement of the total
phosphorus column density along a given sightline but with little information on the
velocity structure. One should note that analyzing this line requires high--quality data
and a good knowledge of FPN for the \textit{FUSE} detectors. Indeed, duplicate detectors
make more easy to discriminate FPN and absorption features in most cases. Unfortunately,
in some cases this line appears slightly blended with the broad Fe\3* $\lambda$1124
stellar line.

\begin{figure}[h!]
\hspace{0.4cm}
 \epsfxsize=6cm
\rotatebox{0}{\epsfbox{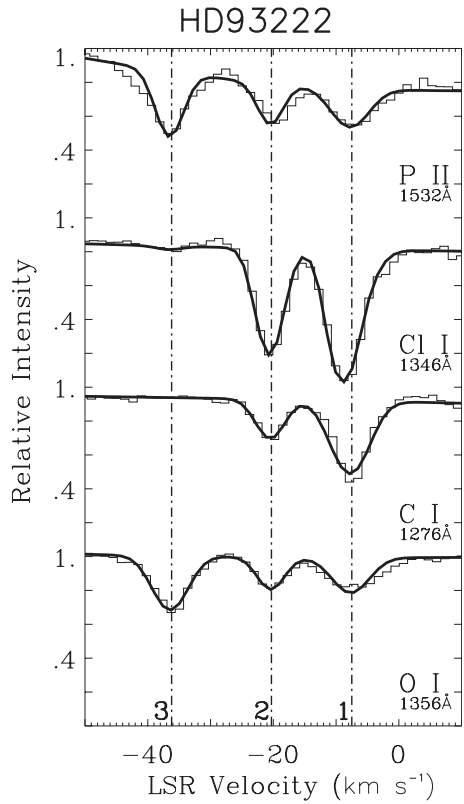}}
  \hspace{0.4cm}
\caption[]{\small{P\2\ $\lambda$1533 absorption line profile toward HD93222. Three
components are easily identified. When detected, Cl\1, C\1, and S\1\ lines are used to
constrain the velocity distribution (turbulent velocity and radial velocity). P\2\ and
O\1\ lines have very similar optical depths since the $N \cdot f$ product is
approximately the same (see text).}} \label{fig:HD93222} \vspace{0.1cm}
\end{figure}

Slightly stronger is the P\2\ line at 1532.5330 \AA\ ($f = 0.303\times10^{-2}$,
observable with \textit{STIS}). Thanks to the \textit{STIS} higher resolution, this line
allows the investigation of the detailed velocity distribution of sightlines and the
derivation of phosphorus column densities in individual clouds. We thus used this line to
derive the P\2\ column densities (column density of each individual clouds and integrated
column density over the sightline) and compare with the O\1\ values. The other great and
main advantage of this line is that its optical depth is systematically found to be
similar to the optical depth of the O\1\ $\lambda$1356 line (see Fig.~\ref{fig:HD93222}).
This is due to the combination of the O\1\ column density, about 2000 times larger than
the P\2\ one, and the oscillator strength, 2000 times lower than for the P\2\
$\lambda$1533 line. Hence, the simultaneous analysis of these two lines minimizes
possible systematic errors due to possible saturation and/or unresolved components. This
combination further allows to investigate individual cloud column densities since these
two lines are observed in the same high-resolution dataset.

\begin{figure}[h!]
\hspace{0.4cm}
 \epsfxsize=8cm
\rotatebox{-90}{\epsfbox{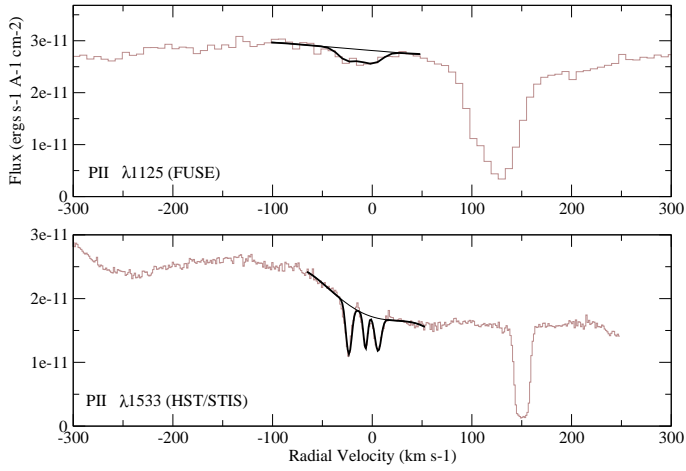}}
  \hspace{0.4cm}
\caption[]{\small{Plot of the P\2\ $\lambda$1125 line toward HD93222 as observed with
\textit{FUSE} (top)  and of the $\lambda$1533 line as observed with \textit{STIS}
(bottom). The two lines have approximately the same oscillator strength. Given the
relatively low spectral resolution, the \textit{FUSE} observation allow to calculate the
integrated column density along the sightline while the \textit{STIS} observation gives
the possibility to investigate column densities of each component.}} \label{fig:PII1124}
\vspace{0.1cm}
\end{figure}

Another possibility consists in comparing the integrated column density as derived from
the unresolved profile of the P\2\ $\lambda$1125 line with \textit{FUSE}, with the sum of
the P\2\ column densities of individual clouds along the sightlines as derived with the
\textit{STIS} resolved profile of the P\2\ $\lambda$1533 line (Fig.~\ref{fig:PII1124}).
We get consistent findings within the error bars (see Table~\ref{tab:glob}). This
comparison provides a twofold confirmation: 1) the relative compatibility of the
corresponding oscillator strengths, and 2) the absence of gross systematic errors when
assuming only one global velocity component to estimate integrated column densities.

\section{Results}\label{sec:results}

\subsection{Abundance of phosphorus}\label{sec:abundance}

\begin{figure}[h!]
\hspace{0.4cm}
 \epsfxsize=6cm
\rotatebox{-90}{\epsfbox{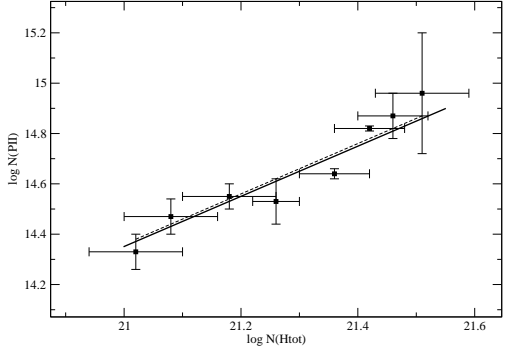}}
  \hspace{0.4cm}
\caption[]{\small{Plot of the P\2\ column density integrated over each sightline versus
the total hydrogen column density defined as $\log [N(\textrm{H\1}) + 2 \times
N(\textrm{H}_2)]$. The thick line shows the solar phosphorus abundance (Asplund et al.
2004) and the thin dashed line shows the regression fit.}} \label{fig:PvsH}
\vspace{0.1cm}
\end{figure}

We plot in Fig.~\ref{fig:PvsH} the integrated P\2\ column density (as derived with the
P\2\ $\lambda1532$ line) versus the total hydrogen content defined as
$N$(H$_\textrm{tot}$)=$N$(H\1)$+2 \times N$(H$_2$). We use the values $N$(H\1) and
$N$(H$_2$) derived by A2003 through \textit{FUSE} observations, available for 8 of our 10
sightlines. Column densities are integrated along sightlines and sample the diffuse ISM.
As compared with the solar abundance $\log$~(P/H)$_\odot$=$-6.65\pm0.04$ (Asplund et al.
2004), the regression of P\2\ vs. H$_\textrm{tot}$ gives a consistent value
$-6.64\pm0.05$. We can conclude that:
\begin{itemize} \item[$\bullet$] Phosphorus is not depleted in the diffuse
neutral ISM up to at least $\log N$(H$_\textrm{tot}$)$\approx 21.5$ and $E$(B-V)$\approx
0.5$,
 \item[$\bullet$] P\1\ exist only in negligible amounts in this gaseous phase since we would observe an
even higher P/H ratio. P\2\ is indeed the dominant state of phosphorus along these 8
sightlines.
\end{itemize}

Of course, these findings must be tempered by the fact that we used the integrated column
densities along the sightlines. As a matter of fact, P\2\ column densities can be derived
for individual clouds (see next section), but unfortunately, it is not possible to do the
same for hydrogen, given the relatively large width of the H\1\ absorption lines observed
with \textit{FUSE} and HST/\textit{STIS}.

\subsection{Phosphorus versus oxygen}\label{sec:parallel}

\begin{figure}[h!]
\hspace{0.4cm}
 \epsfxsize=6cm
\rotatebox{-90}{\epsfbox{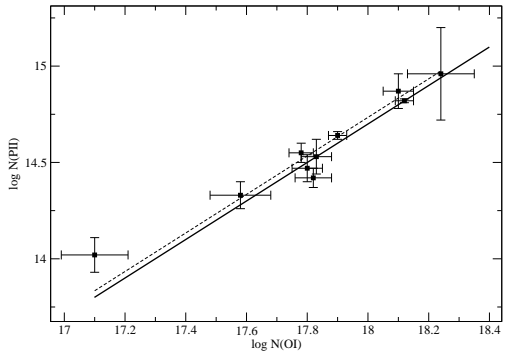}}
  \hspace{0.4cm}
\caption[]{\small{P\2\ total column density over each sightline is plotted versus the
total O\1\ column density. The thick line shows the solar ratio (Asplund et al. 2004) and
the thin dashed line shows the regression fit.}} \label{fig:glob} \vspace{0.1cm}
\end{figure}

The integrated P\2\ and O\1\ column densities toward the 10 Galactic sightlines of our
sample were derived using the P\2\ $\lambda1533$ and O\1\ $\lambda1356$ lines, observable
with \textit{STIS}. Values are listed in Table~\ref{tab:glob} and plotted in
Fig.~\ref{fig:glob}. We find consistent integrated O\1\ column densities within
1~$\sigma$ error bars with those found by A2003 except toward a few sightlines for which
error bars of A2003 could have been somewhat underestimated. There are also some overlap
with other past measurements (see Howk et al. 2000, Jensen et al. 2005, and Knauth et al.
2003). These past values are marginally consistent with our values within 1~$\sigma$
error bars, except the O\1\ column density toward HD218915 derived by Howk et al. (2000).
However, this Howk et al. determination is also inconsistent with the other studies of
A2003 and Knauth et al. 2003.

There is a clear correlation between P\2\ and O\1\ column densities. The derived
error-weighted mean P\2/O\1\ ratio is $-3.26^{+0.12}_{-0.10}$ (1~$\sigma$ error bars),
which is consistent with the solar P/O proportion $-3.30\pm0.07$ (Asplund et al. 2004).
The only exception is the sightline toward the star HD121968 located in the halo. We
observe a single absorption component which could arise in a cloud located in the
low-density partly-ionized Galactic halo. Therefore, in such extreme conditions, P\2\
could also exist where oxygen is ionized so that the actual P/O ratio would be $\eta
\times N$(P\2)/$N$(O\1), where $\eta < 1$ is an unknown factor which depends on the
ionization conditions within the cloud. This ionization effect is negligible for all the
other sightlines.

\begin{figure}[h!]
\hspace{0.4cm}
 \epsfxsize=6cm
\rotatebox{-90}{\epsfbox{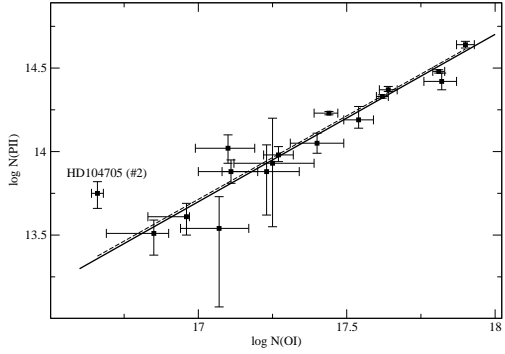}}
  \hspace{0.4cm}
\caption[]{\small{P\2\ column density of each individual component along the sightlines
plotted against the corresponding O\1\ column density. The thick line shows the solar
ratio (Asplund et al. 2004) and the thin dashed line shows the regression fit.}}
\label{fig:ind} \vspace{0.1cm}
\end{figure}

Does the same correlation hold if we plot the column densities of individual clouds along
the sightlines ? We selected individual components in each sightline on the basis of a
clear separation from nearby other absorptions $-$ i.e. the wavelength shift due to the
different radial velocities of clouds must be larger than the intrinsic line widths
(convolution of the turbulent velocity and the instrumental line spread function). We
identified 17 interstellar absorbing regions along our sightlines (Table~\ref{tab:ind}).
We still find a correlation over more than $\approx$1~dex in column density
(Fig.~\ref{fig:ind}), with a mean P\2/O\1\ ratio of $-3.25^{+0.12}_{-0.10}$, again
consistent with the solar P/O ratio. The only noticeable exception to this good
correlation (apart from the single component along the HD121968 sightline, see above) is
the component \#2 along the HD104705 sightline. This sightline crosses an interarm region
in the Milky Way disk where the medium is particularly ionized (Sembach 1994). This
strongly suggests that the absorption component \#2 lies in such a region where P\2/O\1\
is larger than the actual P/O ratio.

Given these correlations, the following conclusions can be put forward:
\begin{itemize}
\item[$\bullet$] There is no clear evidence of a differential depletion of P\2\ and O\1\
in the diffuse neutral gas sampled here. This is true over all the range of molecular
fraction $f$(H$_2$) = 0.05-0.28 and reddening $E$(B-V) = 0.17-0.53. \item[$\bullet$] P\2\
and O\1\ coexist in the same gaseous regions; no ionization correction seems to be
required, except in two ionized interstellar clouds. Phosphorus and oxygen are thus
likely to evolve in parallel in the ISM since the time they are produced in stars and to
be well mixed. \item[$\bullet$] Finally, the oscillator strength of the P\2\
$\lambda$1533 line seems to be well defined and compatible with the $f$-value of the
$\lambda$1125 line (see Sect.~\ref{sec:data}).
\end{itemize}

\subsection{Measurements in various physical environments}\label{sec:others}

We gathered from the literature several Galactic and extragalactic sightlines toward
which both P\2\ and O\1\ column densities have been measured (see Table~\ref{tab:lit}).
We have excluded two sightlines toward subdwarf O stars (Wood et al. 2004) since the P\2\
column densities were determined through only one saturated line.

\begin{figure*}[h!]
\hspace{0.4cm}
 \epsfxsize=12cm
\rotatebox{-90}{\epsfbox{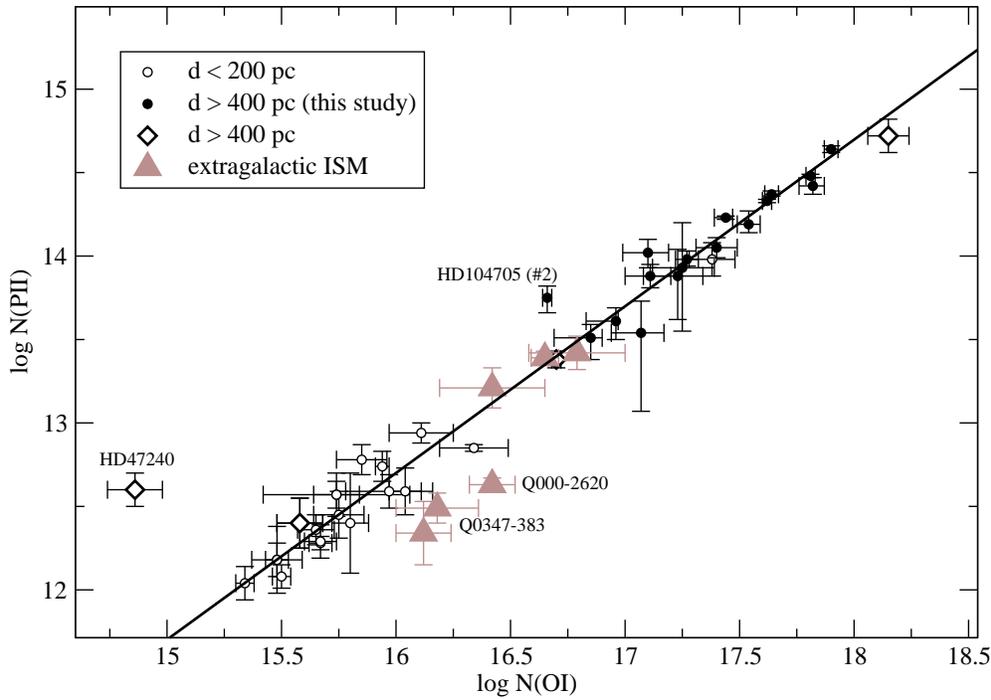}}
  \hspace{0.4cm}
\caption[]{\small{Compiled measurements of P\2\ and O\1\ column densities found in the
literature (see Table~\ref{tab:lit}) overplotted on our results. Empty circles represent
sightlines toward stars with distances less than 200~pc. Filled circles represent the
values of this study (results plotted in Fig.~\ref{fig:ind} toward stars between $\approx
400$ and $\approx 5000$~pc), and diamonds represent values found in the literature toward
stars with comparable distances. Finally, triangles show measures in the extragalactic
ISM. The thick line represent the solar P/O proportion (Asplund et al. 2004).}}
\label{fig:lit} \vspace{0.1cm}
\end{figure*}

The sightlines are quite various in terms of distances to the targets. We list indeed
twenty stars, from the Local Bubble (within 100~pc) to the distant Galactic plane (with
star distances up to 5~kpc), one star in the Small Magellanic Cloud, and four
high-redshift damped Lyman $\alpha$ systems (DLAs). The sightlines also span various
metallicities (from solar to $\sim 1/100$ solar) and various color excesses. All these
values are overplotted on our results in Fig.~\ref{fig:lit}. The correlation between P\2\
and O\1\ column densities found in Sect.~\ref{sec:parallel} still holds over more than
three decades in column densities.

We notice however that in three DLAS, the P\2/O\1\ column density ratio is significantly
lower than the solar P/O value. This could be the sign of a metallicity dependence of
P\2/O\1\ although the measures in other low metallicity environments give solar values
(Table~\ref{tab:lit}). On contrary, Welsh et al. (2001) found a relatively high P\2/O\1\
ratio in the high velocity gas toward HD47240, lying just behind the Monoceros Loop
Supernova Remnant (SNR). This high ratio could be explained by the ionization effects of
the SNR gas found by the authors.

The derived error-weighted mean P\2/O\1\ ratio of all these data points is
$-3.28^{+0.15}_{-0.13}$, again consistent with the solar P/O value. Results of previous
section are thus confirmed and reinforced since P\2/O\1\ appears to be relatively
homogeneous, at least within the Milky Way over nearly 5~kpc.

As a conclusion, it is thus reasonable to propose that phosphorus, as traced by P\2, can
be used to trace O\1\ column density in the diffuse ISM since there is no clear sign of
differential depletion.

\section{Conclusions}

We investigated 10 Galactic sightlines and found that:
\begin{itemize}
\item[$\bullet$] P\2\ is a good tracer of the neutral gas in most interstellar clouds and
can be used as a proxy to derive the phosphorus abundance in this gaseous phase.
Ionization seems to be an issue only in extreme conditions.
 \item[$\bullet$] There is no depletion of phosphorus in the diffuse neutral medium toward these stars.
\item[$\bullet$] P\2\ and O\1\ column densities relate to each other in solar P/O
proportions. There is no differential depletion of P\2\ and O\1. \item[$\bullet$] The
oscillator strengths of the P\2\ $\lambda$1125 and P\2\ $\lambda$1533 lines do not suffer
from significant misestimates.
\end{itemize}

Our results suggest in particular that phosphorus could be an ideal tracer of oxygen in
extragalactic regions. Oxygen is indeed of prime importance to understand the chemical
evolution of galaxies. It is one of the most abundant element in the Universe and its
nucleosynthesis is well known. Its abundance is therefore widely used to estimate the
metallicity of different Galactic and extragalactic ISM gas phases. Oxygen has been in
particular extensively investigated to derive H\2\ region metallicities. However, in the
neutral phase, oxygen abundance in extragalactic regions is often poorly constrained.
Indeed, the O\1\ lines detected in blue compact dwarfs (BCDs) and in giant H\2\ regions
of spiral galaxies are often saturated resulting in large error bars on the metallicity
determinations. Among the BCDs for which the neutral gas have been probed so far with
\textit{FUSE}, several suffer from large systematic errors on their O\1\ column
densities, complicating a further interpretation. The errors on $\log$~(O\1/H\1) are
generally larger than $\pm0.5$~dex (Lecavelier et al. 2003, Thuan et al. 2002, Thuan et
al. 2005, Lebouteiller et al. 2004, Cannon et al. 2003). To circumvent this issue, the
use of phosphorus could be a interesting new way to estimate the oxygen abundance,
provided ionization corrections are not needed. However, note that if the spread of the
P\2/O\1\ ratio is only due to ionization and not to depletion or other effects, it can
reveal a useful constraint to the ionization corrections in diffuse clouds. Finally, he
present study is too limited to answer the question of the metallicity dependence of the
P\2/O\1\ ratio. Therefore, we cannot conclude without ambiguities on the use of
phosphorus in low-metallicity environments such as DLAs.

\begin{acknowledgements}
     This work was done using the code {\it Owens}.f developed by M. Lemoine and
     the {\it FUSE} French team. We wish to thank Guillaume H\'ebrard, Jean-Michel
D\'esert, and Daniel Kunth for their help, and for useful discussions.
\end{acknowledgements}

\begin{table*}[b!]
      \caption[]{Stellar and sightline properties. Distances, reddening, and spectral types are from
Diplas and Savage (1994) unless otherwise noted. Molecular hydrogen fraction along the
sightlines are from A2003. The typical uncertainty in the distance is 30~\%.}
         \label{tab:prop}
         \begin{tabular}{lcccccl}
            \hline\hline
            \noalign{\smallskip}
                 Star       & \textrm{d (pc)} &  \textrm{l}     & \textrm{b} & $E$(B-V)& $f$(H$_2$)  & \textrm{Spectral type}      \\
            \noalign{\smallskip}
            \hline
            \noalign{\smallskip}
HD24534     &  450&163.1 &$-17.1$ &0.59& / & \textrm{O9.5V }\\
HD93222$^a$ & 2900&287.7 & $-1.0$ &0.36& 0.05 & \textrm{O7 III}\\
HD99857     & 3060&295.0 &$-4.9$  &0.33& 0.24 &\textrm{B0.5Ib}\\
HD104705& 3900 & 297.4 & $-0.3$ &0.26&0.16 & \textrm{B0 III/IV}\\
HD121968& 3620  & 334.0 & $+55.8$ &0.07& / & \textrm{B5}\\
HD124314&  1150  & 312.7 & $-0.4$ & 0.53& 0.20 &\textrm{O7} \\
HD177989& 5010& 17.8 & $-12.0$ &0.65&0.25 &\textrm{B2 II}\\
HD202347$^a$& 1300 & 88.2 & $-2.0$ &0.17& 0.17 & \textrm{B1 V}\\
HD218915    & 2480 &109.3 & $-1.8$ &0.17& 0.18 & \textrm{O9.0III}\\
HD224151& 1360& 115.4 & $-4.6$ &0.44& 0.28 &\textrm{B0.5 III}\\
            \hline
         \end{tabular}
\begin{list}{}{}
\item[$^{\mathrm{a}}$] Distances, reddenings, and spectral types from Savage et al.
(1985).
\end{list}
   \end{table*}

\begin{table*}[b!]
\caption{\small{Summary of \textit{FUSE}
observations.}}\label{tab:fuseobs}
\begin{center}
\begin{tabular}{llccll}
\hline \hline
 Target & Program ID & Exp. Time (ksec) & Number of Exp. & Aperture &  Mode \\
\hline
HD24534 & P1930201 & 8.3 & 8 & LWRS &  TTAG \\
HD93222 & P1023701 &  3.9& 4 &LWRS & HIST \\
HD99857 &P1024501 &4.3  & 7 & LWRS & HIST\\
HD104705 &P1025701 &4.5  & 6  &LWRS & HIST\\
HD121968 &P1014501 & 9.2 & 18 & LWRS&HIST \\
HD124314 & P1026201 &4.4  & 6 & LWRS & HIST\\
HD177989 &P1017101 &10.3  & 20  &LWRS & HIST\\
HD202347 & P1028901 &0.1  & 1 &LWRS & HIST\\
HD218915 &P1018801 &5.4  & 10  &LWRS & HIST\\
HD224151 & S3040202 & 13.4 & 4 & LWRS & TTAG \\
 \hline
\end{tabular}
\end{center}
\end{table*}

 \begin{table*}[h!]
      \caption[]{Log of HST/\textit{STIS} observations. All the data have been obtained with the E140H grating.}
         \label{tab:stisobs}
         \begin{tabular}{llcccc}
            \hline\hline
            \noalign{\smallskip}
             Target  &  \textrm{Dataset} &  \textrm{Expo. Time (ksec)} & \textrm{Range (\AA)} & \textrm{Aperture}\\
            \noalign{\smallskip}
            \hline
            \noalign{\smallskip}
HD24534  & \textrm{O66P01020} & 8.8  &  1242-1444 & $0\farcs2\times0\farcs09$  \\
         & \textrm{O66P02010} & 2.0  &  1425-1627 & $0\farcs2\times0\farcs09 $ \\

HD93222  & \textrm{O4QX02010} &  1.7  &  1140-1335 & $0\farcs2\times0\farcs09$  \\
         & \textrm{O4QX02020} &  1.1  &  1315-1517 & $0\farcs2\times0\farcs09 $ \\
         & \textrm{O4QX02030} &  2.5  &  1497-1699 & $0\farcs2\times0\farcs09  $\\

HD99857  & \textrm{O54301010} &  1.3  &  1170-1372 & $0\farcs1\times0\farcs03$  \\
         & \textrm{O6LZ44010} & 1.2  &  1388-1590 & $0\farcs2\times0\farcs2 $ \\

HD104705 & \textrm{O57R01010} &  2.4  &  1170-1372 & $0\farcs2\times0\farcs09$  \\
         & \textrm{O57R01030} &  2.9  &  1388-1590 & $0\farcs2\times0\farcs09 $ \\

HD121968 & \textrm{O57R02010} &  1.6  &  1170-1372 & $0\farcs2\times0\farcs09 $ \\
         & \textrm{O57R02020} &  2.9  &  1170-1372 & $0\farcs2\times0\farcs09 $ \\
         & \textrm{O57R02030} &  8.4  &  1352-1554 & $0\farcs2\times0\farcs09 $ \\

HD124314 & \textrm{O54307010} &  1.5 & 1170-1372 & $0\farcs1\times0\farcs03$  \\
         & \textrm{O54307030} &  1.5 & 1388-1590 & $0\farcs1\times0\farcs03$  \\

HD177989 & \textrm{O57R03020} &  2.9 &   1170-1372 & $0\farcs2\times0\farcs09$  \\
         & \textrm{O57R04020} &  8.7 &  1388-1590 & $0\farcs2\times0\farcs09 $ \\


HD202347 & \textrm{O5G301010} &  0.8  &  1170-1372 & $0\farcs1\times0\farcs03$  \\
         & \textrm{O5G301040} &  0.9  &  1388-1590 & $0\farcs1\times0\farcs03$  \\

HD218915 & \textrm{O57R05010} &  2.0 &  1170-1372 & $0\farcs2\times0\farcs09 $ \\
         & \textrm{O57R05030} &  1.3 &  1388-1590 & $0\farcs1\times0\farcs03 $ \\

HD224151 & \textrm{O54308010} &  1.5 &  1170-1372 & $0\farcs1\times0\farcs03 $ \\
        & \textrm{O6LZ96010}  &  0.3 & 1388-1590 & $0\farcs2\times0\farcs2 $\\

\hline
\end{tabular}
   \end{table*}

\begin{table*}
      \caption[]{Column densities integrated over the sightlines. H\1\ and H$_2$ measurements are from A2003.
      The \textit{FUSE} determination of the P\2\ column density makes use of the $\lambda1124$ line while the
       HST/\textit{STIS} determination makes use of the $\lambda1532$ line. O\1\ is analyzed through the
       $\lambda1356$ transition.}
         \label{tab:glob}
         \begin{tabular}{lccccccc}
            \hline\hline
            \noalign{\smallskip}
Sightline  &  $\log N$\textrm{(H\1)} &   \textrm{$f$(H$_2$)} &    $\log N$\textrm{(P\2)}  &  $\log N$\textrm{(P\2)}  &   $\log N$\textrm{(O\1)}  &  $\log$~\textrm{P\2/O\1}\\
      &                          &                  &     $[$FUSE$]$     &     $[$STIS$]$             &  $[$STIS$]$             &  $[$STIS$]$            &             \\
           \hline
HD24534 &    /     &      /     &     /        & $14.42\pm0.05$ & $17.82\pm0.06$ &  $-3.40\pm0.08$ \\
         &          &        &        &                   &                   (/)$^{\mathrm{a}}$ & &      \\
HD93222& $21.40$ &  $0.05$ &  $14.79\pm0.06$ & $14.82\pm0.01$ & $18.12\pm0.03$ & $-3.30\pm0.03$ \\
         &          &        &        &                   &                   ($18.13\pm0.02$)$^{\mathrm{a}}$ & &      \\
HD99857&  $21.24$ &  $0.24$ &  $14.67\pm0.06$ & $14.64\pm0.02$ & $17.90\pm0.03$ & $-3.26\pm0.03$ \\
         &          &        &        &                   &                   ($17.89\pm0.03$)$^{\mathrm{a}}$ & &      \\
HD104705 & $21.10$ &  $0.16$ &  $14.75 \pm 0.04$ & $14.54 \pm 0.04$ & $17.77\pm 0.04$ &$-3.23 \pm 0.06$ \\
         &          &        &        &                   &                   ($17.81\pm0.02$)$^{\mathrm{a}}$ & &      \\
HD121968 &    /      &     /    &  $14.04\pm0.09$ & $14.02\pm0.09$ & $17.10\pm0.11$ &$-3.08\pm0.15$ \\
         &          &        &        &                   &                   (/)$^{\mathrm{a}}$ & &      \\
HD124314 & $21.41$ &  $0.20$ &  $14.81 \pm 0.04$ & $14.96 \pm 0.24$ & $18.24 \pm 0.11$ &$-3.28 \pm 0.28$ \\
         &          &        &        &                   &                   ($18.18\pm0.02$)$^{\mathrm{a}}$ & &      \\
HD177989& $20.96$ &  $0.25$ &  $14.43 \pm 0.04$ & $14.47 \pm 0.07$ & $17.80 \pm 0.05$ &$-3.33 \pm 0.09$ \\
         &          &        &        &                   &                   ($17.79\pm0.03$)$^{\mathrm{a}}$ & &      \\
HD202347 & $20.94$ &  $0.17$ &  $14.34\pm0.17$ & $14.33\pm0.07$ & $17.58 \pm 0.10$ &$-3.26\pm0.13$ \\
         &          &        &        &                   &                   ($17.58\pm0.06$)$^{\mathrm{a}}$ & &      \\
HD218915 & $21.17$ & $0.18$ &  $14.37\pm0.10$ & $14.53\pm0.09$ & $17.83\pm0.05$ &$-3.30\pm0.11$ \\
         &          &        &        &                   &                   ($17.82\pm0.03$)$^{\mathrm{a}}$ & &      \\
HD224151 & $21.32$ &  $0.28$ & /                  & $14.87\pm0.09$ & $18.10\pm0.05$ & $-3.23\pm0.11$ \\
         &          &        &        &                   &                   ($18.06\pm0.03$)$^{\mathrm{a}}$ & &      \\
            \hline
         \end{tabular}
\begin{list}{}{}
\item[$^{\mathrm{a}}$] When measured, O\1\ column densities derived in A2003.
\end{list}
   \end{table*}

\begin{table*}
      \caption[]{Column densities of the individual clouds along each sightline. No isolated component is
      found in the O\1\ $\lambda$1356 and P\2\ $\lambda$1533 lines for HD224151.}
         \label{tab:ind}
         \begin{tabular}{lccccc}
            \hline
            \hline
            \noalign{\smallskip}
             Sightline  &  Component \textrm{\#} & \textrm{ V$_{\rm LSR}$ } & $\log N$\textrm{(O\1)} & $\log N$\textrm{(P\2)} & $\log \textrm{P\2/O\1}$ \\
                  &       &  \textrm{(km s$^{-1}$)} &    &  &  \\
            \noalign{\smallskip}
            \hline
            \noalign{\smallskip}
HD24534 &  1 & $13.8$  &  $17.82\pm0.06$  &  $14.42\pm0.11$  & $-3.40\pm0.08$   \\
HD93222 &  1 & $-8.9$  &  $17.44\pm0.05$  &  $14.23\pm0.01$  & $-3.21\pm0.05$  \\
        &  2 & $-25.6$ &  $17.81\pm0.02$  &  $14.48\pm0.01$  & $-3.33\pm0.02$   \\
        &  3 & $3.6$   &  $17.64\pm0.03$  &  $14.37\pm0.02$  & $-3.27\pm0.04$    \\
HD99857 &  1 & $6.5$   &  $17.90\pm0.03$  &  $14.64\pm0.02$  & $-3.26\pm0.04$  \\
HD104705&  1 & $2.2$   &  $17.62\pm0.02$  &  $14.33\pm0.01$  & $-3.29\pm0.02$  \\
        &  2 & $-30.4$ &  $16.66\pm0.02$  &  $13.75\pm0.09$  & $-2.91\pm0.09$ \\
        &  3 & $-19.9$ &  $17.11\pm0.11$  &  $13.88\pm0.07$  & $-3.23\pm0.13$ \\
HD121968&  1 & $-10.3$ &  $17.10\pm0.11$  &  $14.02\pm0.09$  & $-3.08\pm0.15$  \\
HD124314&  1 & $0.5$   &  $17.23\pm0.15$  &  $13.88\pm0.22$  & $-3.35\pm0.30$ \\
        &  2 & $-10.3$ &  $17.07\pm0.13$  &  $13.54\pm0.31$  & $-3.53\pm0.37$ \\
        &  3 & $-21.0$ &  $17.25\pm0.14$  &  $13.93\pm0.33$  & $-3.08\pm0.15$ \\
HD177989&  1 & $-2.4$  &  $17.54\pm0.05$  &  $14.19\pm0.07$  & $-3.35\pm0.09$  \\
HD202347&  1 & $-13.8$ &  $17.40\pm0.09$  &  $14.05\pm0.06$  & $-3.35\pm0.11$  \\
        &  2 & $-8.4$  &  $16.96\pm0.13$  &  $13.61\pm0.11$  & $-3.35\pm0.18$  \\
HD218915&  1 & $-9.1$  &  $17.27\pm0.05$  &  $13.98\pm0.05$  & $-3.29\pm0.07$  \\
        &  2 & $-19.5$ &  $16.85\pm0.13$  &  $13.51\pm0.11$  & $-3.34\pm0.18$  \\
            \hline
         \end{tabular}
\begin{list}{}{}
\item[$^{\mathrm{*}}$] Number within parenthesis are 1--$\sigma$ error bars.
\end{list}
   \end{table*}

 \begin{table*}
      \caption[]{Compilation of O\1\ and P\2\ measurements found in the literature.
      The distance we report is the target distance, it is also an upper limit for the cloud distances.}
         \label{tab:lit}
         \begin{tabular}{lcccll}
            \hline
             Sightline  &  $\log N$(O\1) & $\log N$(P\2) & $\log$~P\2/O\1 & reference  & comment\\
                       \hline
WD0004+330 &$16.34\pm0.15$& $12.85\pm0.02$& $-3.50\pm0.15$& \textrm{Lehner et al. 2003} & Star distance $D=97$~pc\\
WD0131$-$163 &$15.85\pm0.11$& $12.78\pm0.09$& $-3.07\pm0.14$&        ''                 & $D=96$ pc\\
WD1211+332 &$15.74\pm0.32$& $12.57\pm0.08$& $-3.17\pm0.32$&        ''                 & $D=115$ pc\\
WD1528$-$163 &$15.80\pm0.08$& $12.40\pm0.30$& $-3.40\pm0.30$&        ''                 & $D=140$ pc\\
WD1631+781 &$15.78\pm0.11$& $12.45\pm0.14$& $-3.33\pm0.17$&        ''                 & $D=67$ pc\\
WD1634$-$573$^{\mathrm{a}}$ &$15.50\pm0.04$& $12.08\pm0.07$& $-3.43\pm0.08$&        ''                 & $D=37$ pc\\
WD1800+685 &$16.11\pm0.14$& $12.94\pm0.06$& $-3.17\pm0.15$&        ''                 & $D=159$ pc\\
WD1844$-$223 &$15.97\pm0.19$& $12.59\pm0.10$& $-3.38\pm0.20$&        ''                 & $D=62$ pc\\
WD2004$-$605 &$15.65\pm0.08$& $12.36\pm0.09$& $-3.29\pm0.11$&        ''                 & $D=58$ pc\\
WD2011+395 &$16.04\pm0.07$& $12.59\pm0.14$& $-3.45\pm0.15$&        ''                 & $D=141$ pc\\
WD2124$-$224 &$15.94\pm0.03$& $12.74\pm0.09$& $-3.20\pm0.10$&        ''                 & $D=224$ pc\\
WD2211$-$495$^{\mathrm{b}}$ &$15.34\pm0.04$& $12.04\pm0.10$& $-3.30\pm0.11$&        ''                 & $D=53$ pc\\
WD2309+105 &$15.67\pm0.05$& $12.28\pm0.04$& $-3.39\pm0.07$&        ''                 & $D=79$ pc\\
WD2331$-$475 &$15.48\pm0.05$& $12.18\pm0.10$& $-3.30\pm0.11$&        ''                 & $D=82$ pc\\
\hline
HD185418 & $18.15\pm0.09$ & $14.72\pm0.10$ & $-3.43\pm0.14$ & \textrm{Sonnentrucker et al. 2003} & $D=790$ pc \\
\hline
GD 246& $15.67\pm0.07$ & $12.29\pm0.10$ & $-3.38\pm0.13$ & \textrm{Oliveira et al. 2003} &  $D=79$ pc \\
WD 2331$-$475 & $15.48\pm0.11$ & $12.18\pm0.20$ & $-3.30\pm0.25$ & ''&  $D=82$ pc \\
HZ 121 & $15.74\pm0.10$  & $12.57\pm0.13$ & $-3.17\pm0.17$ & ''& $D=115$ pc \\
\hline
$\zeta$Pup & $16.70\pm0.20$ & $13.38\pm0.05$ & $-3.32\pm0.21$ & \textrm{Morton et al. 1978} & $D=450$ pc \\
\hline
$\zeta$Oph & $17.38\pm0.10$ & $13.98\pm0.10$ & $-3.40\pm0.14$ & \textrm{Morton et al. 1975} & $D=200$ pc \\
\hline
$\alpha$Vir & $15.58\pm0.10$ & $12.40\pm0.15$ & $-3.18\pm0.20$ & \textrm{York \&\ Kinahan 1979} & $D=88$ pc \\
\hline
HD47240 & $14.86\pm0.02$ & $12.60\pm0.10$ & $-2.26\pm0.10$ & \textrm{Welsh et al. 2001} & $D=1800$ pc \\
\hline
Sk108 & $16.65\pm0.06$ & $13.39\pm0.04$ & $-3.26\pm0.07$ & \textrm{Mallouris et al. 2003} & (SMC star) \\
             &               &                  &              &                           &  Metallicity Z$\sim$Z$_\odot$/4\\
\hline
Q0347$-$383$^{\mathrm{c}}$ & $16.18\pm0.18$ & $12.48\pm0.09$ & $-3.70\pm0.21$ & \textrm{Ledoux et al. 2003} & (DLA) \\
             &               &                  &              &                           &  Redshift z$_\textrm{abs}$=3.02485, \\
                        &                 &               &               &                              &  Z$\sim$Z$_\odot$/10 \\
Q0347$-$383$^{\mathrm{c}}$ & $16.12\pm0.12$ & $12.34\pm0.19$ & $-3.78\pm0.24$ & ''                        & (DLA) z$_\textrm{abs}$=3.02463, \\
                          &               &                &                &                             & Z$\sim$Z$_\odot$/10 \\
\hline
LLIV Arch & $16.42\pm0.23$ & $13.21\pm0.12$ & $-3.21\pm0.28$ & \textrm{Richter et al. 2001} & Z$\sim$Z$_\odot$  \\
\hline QSO HE 2243-6031  & $16.79\pm0.21$ & $13.42\pm0.10$ & $-3.37\pm0.25$ & \textrm{Lopez et al. 2002} & (DLA) z$_\textrm{abs}$=2.33,\\
             &               &                  &              &                           &   Z$\sim$Z$_\odot$/12 \\
            \hline
QSO 0000$-$2620  & $16.42\pm0.10$ & $12.63\pm0.04$ & $-3.79\pm0.11$ & \textrm{Molaro et al. 2001} & (DLA) z$_\textrm{abs}$=3.3901, \\
              &                &                 &                &                            & Z$\sim$Z$_\odot$/100 (Molaro et al. 2000) \\
            \hline
         \end{tabular}
\begin{list}{}{}
\item[$^{\mathrm{a}}$] Wood et al. (2002) obtained a consistent ratio
$\log$~P\2/O\1=$-3.43\pm0.15$ for this sightline. \item[$^{\mathrm{b}}$] H{\' e}brard et
al. (2002) obtained a consistent ratio $\log$~P\2/O\1=$-3.29\pm0.23$ for this sightline.
\item[$^{\mathrm{c}}$] Levshakov et al. (2002) found $-3.63\pm0.03$ towards this QSO. We
choose to quote the measurement of Ledoux et al. (2003) who obtained significantly better
VLT/\textit{UVES} data and had a particular attention on the saturation of O\1\ lines.
\end{list}
   \end{table*}

%
%

\end{document}